# Atomically Thin Resonant Tunnel Diodes built from Synthetic van der Waals Heterostructures

Yu-Chuan Lin,[a] Ram Krishna Ghosh,[b] Rafik Addou,[c] Ning Lu,[c] Sarah M. Eichfeld,[a] Hui Zhu,[c] Ming-Yang Li,[d] Xin Peng,[c] Moon J. Kim,[c] Lain-Jong Li,[e] Robert M. Wallace,[c] Suman Datta,[b] and Joshua A. Robinson[a],*

**Vertical integration of two-dimensional (2D) van der Waals (vdW) materials is predicted to lead to novel electronic and optical properties not found in the constituent layers. Here, we present the direct synthesis of two unique, atomically thin, multi-junction heterostructures by combining graphene with the monolayer (ML) transition-metal dichalocogenides (TMDs): molybdenum disulfide ($MoS_2$), molybdenum diselenide ($MoSe_2$), and tungsten diselenide ($WSe_2$). The realization of $MoS_2$-$WSe_2$-Graphene and $WSe_2$-$MoSe_2$-Graphene heterostructures leads to resonant tunneling in an atomically thin stack with spectrally narrow room temperature negative differential resistance (NDR) characteristics. Density functional theory (DFT) coupled with non-equilibrium Green's function (NEGF) transport model confirms the experimental phenomenon, and provides evidence that the n-type TMD ($MoS_2$, $MoSe_2$) act as barrier layers that electronically confine the p-type TMD ($WSe_2$), leading to resonant tunneling transport of carriers.**

Resonant tunneling of charge carriers between two spatially separated quantum states can lead to a unique current transport phenomenon known as NDR.[1,2] This is a key feature for novel nanoelectronic circuits that utilize bistability and positive feedback, such as novel memories, multi-valued logic, inductor-free compact oscillators, and many other not-yet-realized electronic applications.[3,4] However, realizing spectrally narrow NDR in a RTD at room temperature has been challenging due to carrier scattering related to interfacial imperfections, which are unavoidable in traditional semiconductor heterostructures synthesized using advanced epitaxial growth techniques.[5] Two-dimensional materials,[6,7] with no out-of-plane chemical bonding and pristine interfaces, presents an appealing alternative to traditional semiconductors, and could ultimately eliminate the interfacial imperfections that limit room temperature NDR performance to-date. Since 2004,[6] the overwhelming majority of electronic transport and stacked in 2D materials has been reported using mechanically exfoliated flakes.[8] Recently, there has been a concerted effort to directly synthesize layered TMDs, with powder vaporization[9–11] synthesis paving the way for direct growth of atomically thin structures.[9–14] Beyond monolayer TMDs, vdW heterostructures (heterogeneous stacks of dissimilar atomic layers) have been predicted to lead to novel electronic properties not found in their constituent layers,[15] where their realization has primarily come from mechanical exfoliation and stacking.[16–19] Manual stacking has provided experimental verification of electronic bandgap modulations and strong interlayer coupling,[20] but it can also lead to interface contamination[19] that introduces additional scattering mechanisms and inhibits the NDR. Therefore, a synthetic route to achieve vdW heterostructures with pristine interfaces

[a]Department of Materials Science and Engineering and Center for 2-Dimensional and Layered Materials, The Pennsylvania State University, University Park, Pennsylvania, 16802, United States; [b]Department of Electrical Engineering, The Pennsylvania State University, University Park, Pennsylvania, 16802, United States; [c]Department of Materials Science and Engineering, The University of Texas at Dallas, Richardson, Texas 75080 United States; [d]Institute of Atomic and Molecular Sciences, Academia Sinica, Taipei 10617, Taiwan; [e]Physical Science and Engineering Division, King Abdullah University of Science and Technology, Thuwal, 23955-6900, Saudi Arabia. *email: jrobinson@psu.edu

will be a critical step in the advancement of the field.

Here, we present the direct synthesis of $MoS_2$-$WSe_2$-Graphene and $WSe_2$-$MoSe_2$-Graphene heterostructures employing a combination of oxide powder vaporization and metal-organic chemical vapor deposition (MOCVD). We not only demonstrate that these heterostructures exhibit the same interlayer electronic coupling found in mechanically exfoliated heterostructures,[20–22] but also show that they exhibit unique electronic transport properties not typically found in exfoliated structures. We discover that direct grown heterostructures exhibit resonant tunneling of charge carriers, which leads to sharp negative differential resistance (NDR) at room temperature. Using a combination of density functional theory (DFT) and non-equilibrium Green's function (NEGF) formalism, we find that the p-type TMD layer ($WSe_2$) acts as the quantum well, where the n-type TMD layer ($MoS_2$, $MoSe_2$) and the graphene layer adjacent to the TMD layers act as tunnel barriers setting up resonant transmission channels at specific energy levels and k-points in the Brillouin region. Importantly, we identify that the peak of the resonant tunneling can be tuned by modifying the stacking order or layer composition, which will be a powerful tool toward engineering heterostructures for ultra-low power electronic devices.

**Result and Discussion**

The formation of ML vdW heterostructures is achieved by sequentially growing two dissimilar TMD monolayers on multi-layer (3 layers) epitaxial graphene (EG) (Fig. 1a).[23] The individual TMD layers are grown *ex-situ* via powder vaporization or MOCVD. Tungsten Diselenide is synthesized using both routes: tungsten trioxide ($WO_3$) and selenium (Se) powders for the powder vaporization route,[24] and tungsten hexacarbonyl ($W(CO)_6$) and dimethylselenium ($(CH_4)_2Se$) for the MOCVD route.[25] Molybdenum disulfide is grown via vaporization of molybdenum trioxide ($MoO_3$) and sulfur.[10] The heterostructure synthesis process is summarized in Fig. 1. The first TMD layer of the heterostructure, $WSe_2$ or $MoS_2$, is grown on tri-layer EG (Fig. 1a) at 950 $^{\circ}$C and 750 $^{\circ}$C for $WSe_2$-EG (Fig. 1b) and $MoS_2$-EG (Figs. 1c and 1d), respectively. Following this first TMD growth step, the surface coverage of the $WSe_2$ or $MoS_2$ on EG is typically >60%, with a lateral size of 2 μm and 300 nm for $WSe_2$ and $MoS_2$, respectively. Subsequently, the $MoS_2$-$WSe_2$-EG vertical heterostructure is created via a second *ex-situ* growth of $MoS_2$ on $WSe_2$-EG at 750 $^{\circ}$C (Fig 1c). Similar to our previous work,[26] we find that wrinkles in the graphene as well as defects and edges within the $WSe_2$ promote vertical growth of the $MoS_2$, and monolayer $MoS_2$/$WSe_2$ is primarily achieved in pristine regions of $WSe_2$ (Fig. 1c and supporting information, Fig. S1).[26] The formation of the $WSe_2$-$MoSe_2$-EG heterojunction occurs during growth of $WSe_2$ on $MoS_2$. During the synthesis, a selenium-sulfur ion exchange occurs when the $MoS_2$ is exposed to the selenium vapor just prior to the growth of $WSe_2$ at 1000 $^{\circ}$C for 45 minutes.[27] Standard topographic characterization via atomic force microscopy (AFM) cannot clearly identify the location of the heterostructures (Fig. 1f), however conductive AFM (CAFM) with platinum ($Pt$) tip[28] provides a means to map the $WSe_2$-$MoSe_2$-EG junctions and adjacent $WSe_2$-EG regions due to a difference in heterostructure conductivity (Fig. 1f and supporting information, Fig. S1).

Raman spectroscopy and transmission electron microscopy (TEM) confirm the formation of crystalline, vertical heterostructures (Figs. 1d-h and supplemental information, Figs. S1-S3). A large fraction of the epitaxial graphene remains nearly defect free following the sequence of TMD growths; however, there are regions of increased defectiveness due to either partial-passivation of the graphene/SiC buffer layer[23] or formation of thick TMD layers.[26] Raman spectroscopy (see supplemental information, Fig. S3) also confirms presence of significant fractions of monolayer $WSe_2$ ($E_{2g}$/$A_{1g}$ at 250 cm$^{-1}$ and 2LA at 263 cm$^{-1}$)[24] and $MoS_2$ ($E_{2g}$ at 383 cm$^{-1}$ and $A_{1g}$ at 404 cm$^{-1}$),[26] as well as monolayer $MoSe_2$ ($A_{1g}$ at 240 cm$^{-1}$ and $E^1_{2g}$ at 284 cm$^{-1}$).[27] X-ray photoelectron spectroscopy (see supplemental information, Fig. S4 and Table S1) also corroborates the absence of any interaction between the two transition metal dichalcogenides or graphene; and indicates that that the $MoS_2$ exhibits an n-type behavior, while the $WSe_2$ layer shows a p-type behavior. Scanning transmission electron microscopy (STEM) (Figs. 1g and Fig. 1h) also verifies the heterostructure is not a manifestation of the alloying of the constituent TMDs, but indeed are unique layers with pristine interfaces with atomic precision. In the case of $MoS_2$-$WSe_2$-EG, we have focused on a multilayer region of $MoS_2$-$WSe_2$ to ensure pristine layer formation beyond the monolayer structure (see supporting information, Fig. S2), however, all electrical characterization presented later is on monolayer heterostructures. The clean interface between monolayers can be observed easily using high resolution STEM. The $WSe_2$-$MoSe_2$-EG ordering is confirmed by comparing the intensity with that of bilayer-$WSe_2$-EG due to the similar atomic number between W and Mo atom (see supplemental information, Fig. S2). Unlike vertical heterostructures based on a single chalcogen (i.e. $MoS_2$/$WS_2$),[29] the ordered layering does not occur when we attempt to grow a vertical structure based on heterogeneous layers where $M_1 \neq M_2$ and $X_1 \neq X_2$ (M = Mo, W; X = S, Se) on "3D" substrates such as sapphire or $SiO_2$ (see supplemental information, Fig. S5). Instead, all attempts to grow such a structure results in alloying or lateral heterostructures of the layers. Therefore we hypothesize that epitaxial graphene plays a critical role in the formation of atomically precise vdW heterostructures where $M_1 \neq M_2$ and $X_1 \neq X_2$ by providing an atomically smooth surface that is free of dangling bonds, enabling mobility on the surface for TMD layer growth. Sapphire and $SiO_2$ surfaces exhibit high surface roughness, dangling bonds, and are therefore more likely to impede surface diffusion, which catalyzes the alloying process.

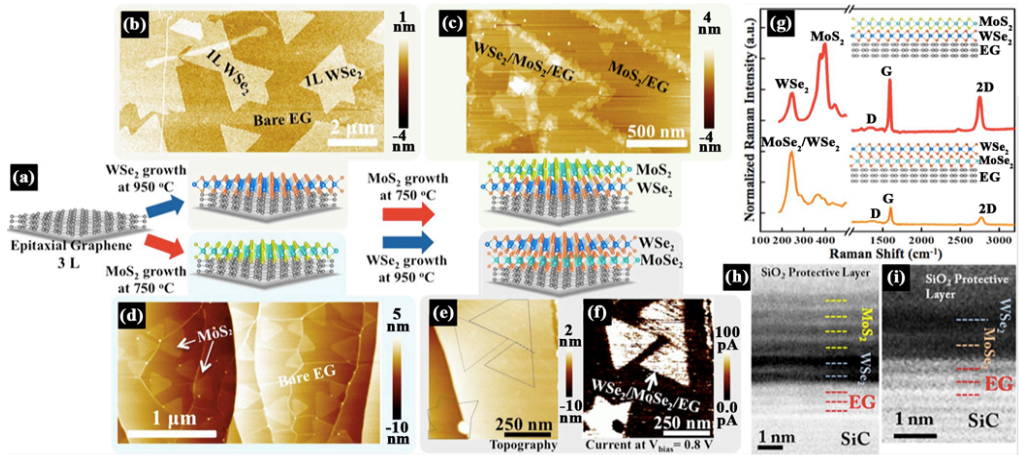

**Figure 1: The formation of vdW Heterostructures.** $MoS_2$-$WSe_2$-EG vertical heterostructures begins with the synthesis of (a) 3L EG from SiC followed by (b) vapor transport or MOCVD of $WSe_2$ and (c) vapor transport of $MoS_2$. $WSe_2$-$MoSe_2$-EG heterostructures are similarly grown, except when (d) $MoS_2$ is grown first on EG followed by (e) growth of the $WSe_2$, a Se-S ion exchange occurs, leading to the formation of $MoSe_2$ from the original $MoS_2$ layer. The $MoSe_2$ domains are difficult to topographically identify; however, (f) conductive AFM clearly delineates their location due to enhanced tunneling at the heterostructures. Raman (g) indicates preservation of the graphene has occurred during the synthesis process, and Scanning TEM (h, i) confirms that the stacked structures exhibit pristine interfaces, with no intermixing of Mo-W or S-Se after synthesis.

Monolayer semiconducting TMDs exhibit a direct optical band gap ($E_{opt}$) ($MoS_2$ at 1.8 ~ 1.9 eV, $MoSe_2$ at 1.55 eV, and $WSe_2$ at 1.6 ~ 1.65 eV);[30] therefore, photoluminescence (PL) spectroscopy (Figs. 2a,b) can provide evidence of electronic coupling between the layers. In addition to the typical PL peaks from the direct bandgap transition within the individual layers, the PL spectra of the heterostructures exhibit the presence of interlayer excitons at 1.59 eV for $MoS_2$-$WSe_2$-EG (Figs. 2a,b) and 1.36 eV for $WSe_2$-$MoSe_2$-EG (Figs. 2a,b). In this case, the $MoS_2$-$WSe_2$ and $WSe_2$-$MoSe_2$ junctions exhibit type II band alignment,[15,20,21,31] where electrons in the $WSe_2$ conduction band transfer to the conduction band of $MoS_2$ ($MoSe_2$) and the excited holes in $MoS_2$ ($MoSe_2$) valence band transfer to the valence band of $WSe_2$. Consistent with manually-stacked heterojunctions,[20,21] the position of the PL peak is due to interlayer exciton recombination, which confirms the electronic coupling at the heterojunction between the two ML TMDs.

Additional evidence of coupling comes from the topographical information of the heterostructures. Similar to graphene-hBN heterostructures,[32] Moiré patterns of $MoS_2$-$WSe_2$ are observed in tapping-mode AFM, which are qualitatively consistent with rotation angles of approximately 0 or 180° between $MoS_2$ and $WSe_2$. Furthermore, scanning tunneling microscopy/spectroscopy (STM/S) (Fig. 2d) confirms the presence of a Moiré pattern produced by the misorientation of $MoS_2$ relative to the underlying WSe$_2$ layer. The lattice constant of the Moiré pattern is 9.8 ± 0.4 nm, which corresponds to a misorientation angle of ~1.9°. Modeling the heterostructure with this misorientation produces a consistent Moiré pattern, with a slightly smaller lattice constant of 9.6 nm (Fig. 2e). While the mechanical stacking technique leads to a variety of rotation angles between layers,[20] the direct growth of vdW layers using our approach appears to have a strict rotational alignment, which may be critical for achieving optimal coupling between the layers.[33,34]

Scanning tunneling spectroscopy further provides evidence that the quasi-particle band gap of MoS$_2$-WSe$_2$-EG is significantly smaller than its WSe$_2$-EG counterpart (Figs. 2f,g, and supplemental information, Fig. S6). Based on STS, we infer that, for WSe$_2$-EG, the conduction band minimum (CBM) is at a sample bias of + 0.71±0.08 V and the valence band maximum (VBM) is at -1.11±0.02 V (green curve in Fig. 2e). This indicates that the quasi-particle band gap ($E_g$) of WSe$_2$ is 1.83±0.05 eV, which is higher than $E_{opt}$ (1.63 eV) due to the large excitonic binding energy in 2D TMDs.[14,22,31,35] On the other hand, MoS$_2$-WSe$_2$-EG exhibits a CBM at +0.34±0.03 V and VBM at -1.31±0.03 V, indicating a quasi-particle interlayer $E_g$ of 1.65 eV±0.02 V, which is slightly larger than its interlayer $E_{opt}$ at 1.59 eV (Fig. 2b) but smaller than the $E_{opt}$ in 1L MoS$_2$-EG.[22,31] Mapping the tunnel current density of WSe$_2$-EG and WSe$_2$-MoSe$_2$-EG heterostructures via conductive AFM[28,36] (Fig. 1f and supplemental information, Fig. S1) provides strong evidence that tunneling is much more readily achieved in WSe$_2$-MoSe$_2$-EG at a tip bias of 1.5 V, indicating a smaller, resonance tunneling, or both may be occurring. Finally, we note that defects, such as grain boundaries and vacancies disrupt the continuity of the Moiré pattern, further emphasizing that imperfections in layers or the interface will significantly impact the electronic behavior of vdW heterostructures (Fig. 2d).

Current-voltage measurements through the heterostructure (carried out via CAFM at room temperature) do not exhibit the traditional p-n junction diode-like transport found in mechanically exfoliated dichalcogenide structures or direct grown single-junction (i.e. WSe$_2$-EG) structures.[20,26,37] Instead, we find that, following a "soft" turn-on, the current exhibits a peak at a certain bias voltage ($V_{peak}$ = + 1.1 and + 0.7 for MoS$_2$-WSe$_2$-EG and WSe$_2$-MoSe$_2$-EG, respectively), then decreases to a minimum before undergoing a "hard" turn on with exponential current increase. This NDR phenomenon only occurs in double junction heterostructures and never in single junction dichalcogenide heterostructures.[26,38] In this work, the peak to valley current ratio (PVCR) is 1.9 for MoS$_2$-WSe$_2$-EG and 2.2 for WSe$_2$-MoSe$_2$-EG (Fig. 3a and supplemental information, Fig. S7-S8, Table S2). NDR has been investigated in traditional semiconductor heterojunctions but never at the monolayer limit,[1–5] and has only very recently been demonstrated in manually stacked heterostructures.[39,40]

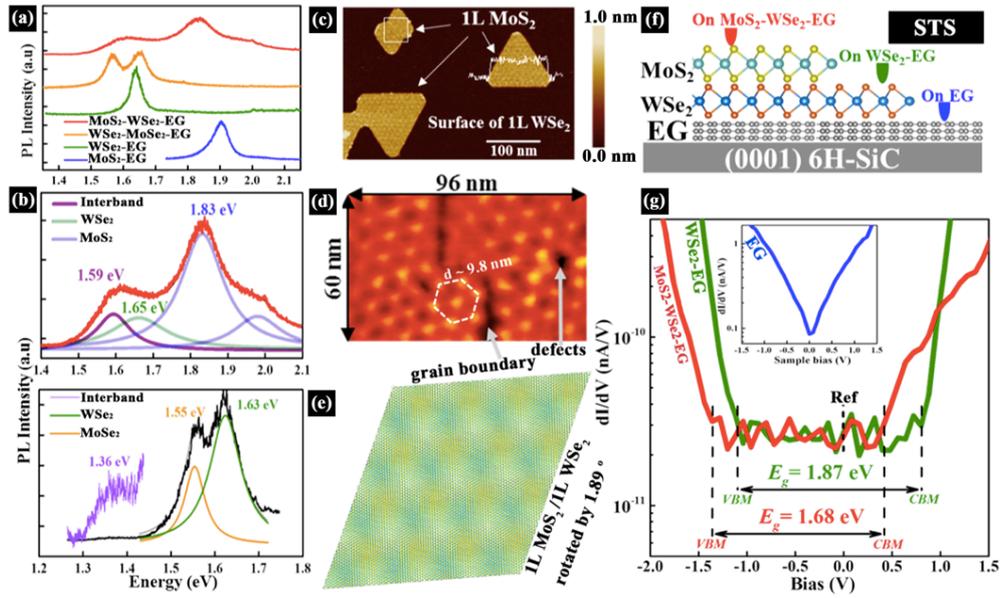

**Figure 2: Coupling in 2D vertical heterostructures.** (a) The PL properties of $MoS_2$-$WSe_2$-EG and $WSe_2$-$MoSe_2$-EG reveal significant interlayer coupling, where the (b) $MoS_2$-$WSe_2$-EG and $WSe_2$-$MoSe_2$-EG exhibit the intrinsic PL peaks corresponding to $MoS_2$, $MoSe_2$, and $WSe_2$, and also exhibit interband PL peaks at 1.59 and 1.36 eV, where the excitation wavelength (λ) is 488 nm and 633 nm for $MoS_2$-$WSe_2$-EG and $WSe_2$-$MoSe_2$-EG, respectively. (c) The moiré patterns acquired via AFM in $MoS_2$ on $WSe_2$ indicates an alignment of nearly either 0° or 180° between the top and bottom layer, and (d) STM confirms the moiré pattern with a lattice constant equal to (9.8 ± 0.4) nm. This structure can be reproduced theoretically (e) when the misorientation angle between these layers is ~1.9°. The continuity of the Moiré pattern is interrupted by the formation of a grain boundary and point defects, as indicated in the STM image. (h) STS on $MoS_2$-$WSe_2$-EG, $WSe_2$-EG, and EG (h, inset) provide evidence that the bandgap of the double junction heterostructure ($MoS_2$-$WSe_2$-EG) is smaller than that of the single junction ($WSe_2$-EG) heterostructure. The positions of CBM, VBM, and quasi-particle bandgap $E_g$ of $WSe_2$ on EG and bilayer on EG are marked.

We perform non-equilibrium ballistic quantum transport calculations by combining density functional theory (DFT) with the non-equilibrium Green's function (NEGF) formalism (details in supplemental information) that provide theoretical *I-V* curves to confirm the NDR transport mechanism in the heterostructure (Figs. 3b,c). In the experimental setup the voltage, $V_{ds}$, is applied between the *Pt* tip of the conducting AFM and the graphene electrode which is grounded, as shown in Fig. 3a. The area of the *Pt* tip is approximately 100 nm$^2$, which in the simulation is modeled as a bulk electrode in the theoretical structure (Fig. 3b, supplemental information, Table S3 and Fig. S9). The calculation produces the bias and the transverse momentum dependent transmission probability (Figs. 3d and 3e) of the carriers tunneling through the heterostructure and is used to simulate the *I-V* characteristics using

Landauer transport formulation:[41]

$$I(V_{ds}) = \frac{2q}{h} \int_{BZ} dk_{||} \int dE \, T(E, k_{||}, V_{ds}) \left[ f\left(\frac{E-E_{f_1}}{k_B T}\right) - f\left(\frac{E-E_{f_2}}{k_B T}\right) \right] \quad (Eq.1)$$

where $E_{f1}-E_{f2}=qV_{ds}$ represents the Fermi window; $BZ$ represents the Brillouin zone; and $T(E,k_{||},V_{ds})$ is the total transmission over the energy channels within the Fermi window calculated self-consistently for each applied bias, $V_{ds}$. Ballistic quantum transport calculations reproduce the NDR in the simulated *I-V* characteristics for $MoS_2$-$WSe_2$-EG heterostructure, in both the positive and negative bias regime (Fig. 3c). The NDR can be explained by revisiting the transmission spectra in Figs. 3d and 3e. The rise and fall of the peaks (identified as P1, P2, P3 for $MoS_2$-$WSe_2$-EG, and P for $WSe_2$-$MoS_2$-EG under forward bias) in the transmission spectra as a function of bias is dependent on resonant tunneling of carriers through the structure, which ultimately leads to the experimentally observed NDR (see supplemental Information, Figs. S10-S12). The resonance transmission peaks in the positive bias regime occurs from the bound hole states in the p-type $WSe_2$ layer which arises from the valence band offset between the n-type $MoS_2$ ($MoSe_2$) and p-type $WSe_2$. Conversely, the bound electron states in the n-type $MoS_2$ ($MoSe_2$) arising from the conduction band offset contribute to the resonance peaks in the negative bias regime of the heterostructure. The theoretical *I-V* traces (Fig. 3c) are in good agreement with the experimental observations (Fig. 3a), providing strong evidence that resonant tunneling is the dominant transport mechanism in the vertical heterostructures.

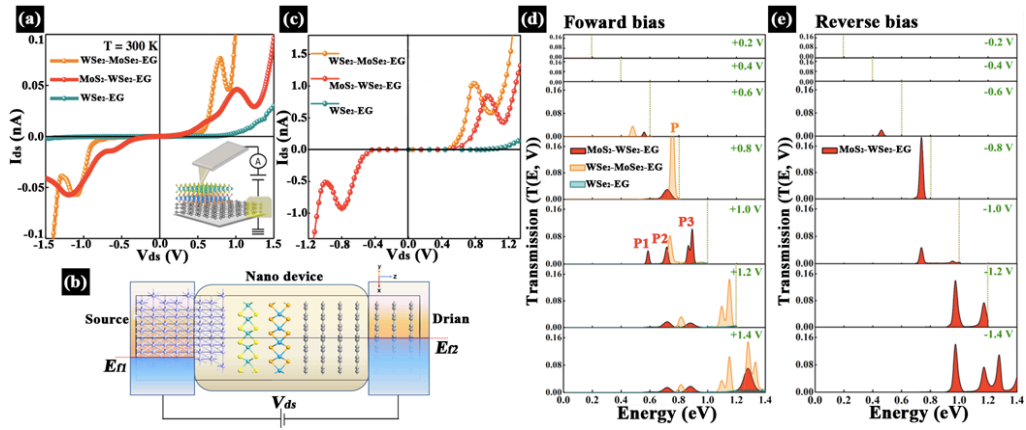

**Figure 3. Resonant tunneling and negative differential resistance in atomically thin layers.** (a) Experimental *I-V* traces for different combination of dichalcogenide-graphene interfaces demonstrating NDR. The inset shows schematic of the experimental setup for the *I-V* measurement in this layered system, (b) schematic of the nano device setup of both of $MoS_2$-$WSe_2$-EG and $WSe_2$-$MoSe_2$-EG system used for non-equilibrium quantum transport calculations by density functional theory (DFT) and non-equilibrium Green's function (NEGF) transport formalism. The theoretical *I-V* curve in (c) is simulated by the DFT and NEGF transport formalism that give (d) – (e) resonant transmission (tunneling) at specific energies and bias voltages. $E_{f1}$ and $E_{f2}$ in (b) indicate the corresponding Fermi levels of the left and right electrodes, respectively, for an applied positive bias $V_{ds}$. This resonant tunneling gives rise to theoretical I-V curves that provides good agreement with the experiment *I-V* characteristics.

## Conclusions

We have demonstrated the direct synthesis of unique multi-junction heterostructures based on graphene (epitaxial graphene on SiC), molybdenum disulfide ($MoS_2$), molybdenum diselenide ($MoSe_2$), and tungsten disulfide ($WSe_2$) that yields pristine interlayer gaps and leads to the first demonstration of resonant tunneling in a atomically thin synthetic stack with the spectrally narrowest room temperature negative differential resistance (NDR) characteristics. This resonant tunneling phenomenon is not typically reported in mechanically stacked dichalcogenide layers even though strong optical coupling between the layers occurs.[20,42] This is due to resonant tunneling being highly sensitive to interfacial perturbations such as defects or "residue" from the transfer process, emphasizing the importance of direct synthesis of multi-junction TMD heterostructures for vertical quantum electronics applications.

## Acknowledgements

Support is acknowledged by the Center for Low Energy Systems Technology (LEAST), one of six centers supported by the STARnet phase of the Focus Center Research Program (FCRP), a Semiconductor Research Corporation program sponsored by MARCO and DARPA. Work at UT-Dallas was also supported by the Southwest Academy on Nanoelectronics (SWAN) a SRC center sponsored by the Nanoelectronics Research Initiative and NIST.

## Contributions

J.A.R. and Y.-C.L. conceived the idea and J.A.R., S.D., R.M.W., M.K. and L.-J.L. directed the research. Y.-C.L., M.-Y.L. and S.M.E. synthesized the heterostructures. Y.-C.L. carried out AFM, Raman, photoluminescence, and C-AFM measurements. R.A. carried out STM/STS, H.Z. and X.P. carried out XPS, N.L. carried out TEM, and R.K.G. carried out the modeling. All authors participated in the analysis of the data and discussed the results. Y.-C.L. and J.A.R. wrote the paper with significant input from all authors. All authors have read and have approved the manuscript.

## Competing financial interests

The authors declare no competing financial interests.

## Reference:

1.	Esaki, L. New Phenomenon in Narrow Germanium p-n Junctions. *Phys. Rev.* **109,** 603 (1958).

2. Esaki, L. & Tsu, R. Superlattice and Negative Differential Conductivity in Semiconductors. *IBM J. Res. Dev.* **14,** 61–65 (1970).

3. Mitin, V. V., Kochelap, V. & Stroscio, M. A. *Quantum Heterostructures: Microelectronics and Optoelectronics*. (Cambridge University Press, 1999).

4. Chan, H. L., Mohan, S., Mazumder, P. & Haddad, G. I. Compact Multiple-valued Multiplexers Using Negative Differential Resistance Devices. *IEEE J. Solid-State Circuits* **31,** 1151–1156 (1996).

5. Bayram, C., Vashaei, Z. & Razeghi, M. AlN/GaN double-barrier resonant tunneling diodes grown by metal-organic chemical vapor deposition. *Appl. Phys. Lett.* **96,** 042103 (2010).

6. Novoselov, K. S. *et al.* Electric field effect in atomically thin carbon films. *Science* **306,** 666–669 (2004).

7. Novoselov, K. S. *et al.* Two-dimensional atomic crystals. *Proc. Natl. Acad. Sci. U. S. A.* **102,** 10451–10453 (2005).

8. Geim, A. K. & Grigorieva, I. V. Van der Waals heterostructures. *Nature* **499,** 419–425 (2013).

9. Zhan, Y., Liu, Z., Najmaei, S., Ajayan, P. M. & Lou, J. Large-area vapor-phase growth and characterization of $MoS_2$ atomic layers on a $SiO_2$ substrate. *Small* **8,** 966–971 (2012).

10. Lee, Y.-H. *et al.* Synthesis of large-area $MoS_2$ atomic layers with chemical vapor deposition. *Adv. Mater.* **24,** 2320–2325 (2012).

11. Gutiérrez, H. R. *et al.* Extraordinary room-temperature photoluminescence in triangular $WS_2$ monolayers. *Nano Lett.* **13,** 3447–3454 (2013).

12. Liu, K.-K. *et al.* Growth of large-area and highly crystalline $MoS_2$ thin layers on insulating substrates. *Nano Lett.* **12,** 1538–1544 (2012).

13. Zhang, Y. *et al.* Direct observation of the transition from indirect to direct bandgap in atomically thin epitaxial $MoSe_2$. *Nat. Nanotechnol.* **9,** 111–115 (2014).

14. Ugeda, M. M. *et al.* Giant bandgap renormalization and excitonic effects in a monolayer transition metal dichalcogenide semiconductor. *Nat. Mater.* **13,** 1091–1095 (2014).

15. Terrones, H., López-Urías, F. & Terrones, M. Novel hetero-layered materials with tunable direct band gaps by sandwiching different metal disulfides and diselenides. *Sci. Rep.* **3,** 1549 (2013).

16. Gao, G. *et al.* Artificially stacked atomic layers: toward new van der Waals solids. *Nano Lett.* **12,** 3518–3525 (2012).

17. Roy, T. *et al.* Field-effect transistors built from all two-dimensional material components. *ACS Nano* **8,** 6259–6264 (2014).

18. Britnell, L. *et al.* Field-effect tunneling transistor based on vertical graphene heterostructures. *Science* **335,** 947–50 (2012).

19. Haigh, S. J. *et al.* Cross-sectional imaging of individual layers and buried interfaces of graphene-based heterostructures and superlattices. *Nat. Mater.* **11,** 764–767 (2012).

20. Fang, H. *et al.* Strong interlayer coupling in van der Waals heterostructures built from single-layer chalcogenides. *Proc. Natl. Acad. Sci. U. S. A.* **111,** 6198–6202 (2014).

21. Rivera, P. *et al.* Observation of Long-Lived Interlayer Excitons in Monolayer $MoSe_2$-$WSe_2$ Heterostructures. *Nat. Commun.* (2015). doi:10.1038/ncomms7242

22. Chiu, M.-H. *et al.* Spectroscopic signatures for interlayer coupling in $MoS_2$-$WSe_2$ van der Waals stacking. *ACS Nano* **8,** 9649–9656 (2014).

23. Robinson, J. A. *et al.* Epitaxial Graphene Transistors: Enhancing Performance via Hydrogen Intercalation. *Nano Lett.* **11,** 3875–3880 (2011).

24. Huang, J.-K. *et al.* Large-area synthesis of highly crystalline $WSe_2$ monolayers and device applications. *ACS Nano* **8,** 923–930 (2014).

25. Eichfeld, S. M., Eichfeld, C. M., Lin, Y.-C., Hossain, L. & Robinson, J. A. Rapid, non-destructive evaluation of ultrathin $WSe_2$ using spectroscopic ellipsometry. *APL Mater.* **2,** 092508 (2014).

26. Lin, Y.-C. *et al.* Direct Synthesis of Van der Waals Solids. *ACS Nano* **8,** 3715–3723 (2014).

27. Su, S.-H. *et al.* Band gap-tunable molybdenum sulfide selenide monolayer alloy. *Small* **10,** 2589–2594 (2014).

28. Lee, G. H. *et al.* Electron tunneling through atomically flat and ultrathin hexagonal boron nitride. *Appl. Phys. Lett.* **99,** 243114 (2011).

29. Gong, Y. *et al.* Vertical and in-plane heterostructures from $WS_2/MoS_2$ monolayers. *Nat. Mater.* **13,** 1135–1142 (2014).

30. Wang, Q. H., Kalantar-Zadeh, K., Kis, A., Coleman, J. N. & Strano, M. S. Electronics and optoelectronics of two-dimensional transition metal dichalcogenides. *Nat. Nanotechnol.* **7,** 699–712 (2012).

31. Chiu, M. *et al.* Determination of band alignment in transition metal dichalcogenides heterojunctions. (2014). arXiv:1406.5137

32. Yang, W. *et al.* Epitaxial growth of single-domain graphene on hexagonal boron nitride. *Nat. Mater.* **12,** 792–797 (2013).

33. Parkinson, B. A., Ohuchi, F. S., Ueno, K. & Koma, A. Periodic lattice distortions as a result of lattice mismatch in epitaxial films of two-dimensional materials. *Appl. Phys. Lett.* **58,** 472 (1991).

34. Klein, A., Tiefenbacher, S., Eyert, V., Pettenkofer, C. & Jaegermann, W. Electronic band structure of single-crystal and single-layer $WS_2$: Influence of interlayer van der Waals interactions. *Phys. Rev. B* **64,** 205416 (2001).

35. Zhang, C., Johnson, A., Hsu, C.-L., Li, L.-J. & Shih, C.-K. Direct imaging of band profile in single layer $MoS_2$ on graphite: quasiparticle energy gap, metallic edge states, and edge band bending. *Nano Lett.* **14,** 2443–2447 (2014).

36. Rawlett, A. M. *et al.* Electrical measurements of a dithiolated electronic molecule via conducting atomic force microscopy. *Appl. Phys. Lett.* **81,** 3043 (2002).

37. Georgiou, T. *et al.* Vertical field-effect transistor based on graphene-$WS_2$ heterostructures for flexible and transparent electronics. *Nat. Nanotechnol.* **8,** 100–3 (2013).

38. Lin, Y.-C. *et al.* Atomically thin heterostructures based on single-layer tungsten diselenide and graphene. *Nano Lett.* **14,** 6936–6941 (2014).

39. Britnell, L. *et al.* Resonant tunnelling and negative differential conductance in graphene transistors. *Nat. Commun.* **4,** 1794 (2013).

40. Roy, T. *et al.* Dual-Gated $MoS_2$/$WSe_2$ van der Waals Tunnel Diodes and Transistors. *ACS Nano* **9,** 2071–2079 (2015).

41. Datta, S. *Quantum Transport-Atom to transistor*. (Cambridge University Press, 2005).

42. Lee, C.-H. *et al.* Atomically thin p–n junctions with van der Waals heterointerfaces. *Nat. Nanotechnol.* **9,** 676–681 (2014).

Supporting Information

# Atomically Thin Resonant Tunnel Diodes built from Synthetic van der Waals Heterostructures

Yu-Chuan Lin,[a] Ram Krishna Ghosh,[b] Rafik Addou,[c] Ning Lu,[c] Sarah M. Eichfeld,[a] Hui Zhu,[c] Ming-Yang Li,[d] Xin Peng,[c] Moon J. Kim,[c] Lain-Jong Li,[e] Robert M. Wallace,[c] Suman Datta,[b] and Joshua A. Robinson[a,*]

**Characterization Instrumentation**

The as-grown heterostructures are characterized using Raman spectroscopy, atomic force microscopy/conductive atomic force microscopy (AFM/CAFM), X-ray photoelectron spectroscopy (XPS), and transmission electron microscopy (TEM). A WITec CRM200 Confocal Raman microscope with 488/514/633 nm wavelength lasers was utilized for structural characterization. A BRUKER Dimension with a scan rate of 0.5 Hz was utilized for the topography image during the AFM measurement. Conductive AFM (CAFM) measurement was performed in PeakForce TUNA mode with platinum (*Pt*) AFM tip. The applied voltage from tips to sample was increased from 0 to 2V. The optimized loading force of the AFM tip and sensitivity was nominally 5 nN and 20 pA/V, respectively, for the I-V measurements carried out on the novel junctions. All the AFM/CAFM measurements in BRUKER Dimension were at room temperature and in ambient. TEM cross-sectional samples were made via utilizing a NanoLab dual-beam FIB/SEM system. Protective layers of $SiO_2$ and Pt were deposited to protect the interesting region during focused ion beam milling. TEM imaging was performed using a JEOL 2100F operated at 200 kV. For surface analysis, the sample was loaded into an ultra-high vacuum (UHV) with a base pressure lower than $2 \times 10^{-10}$ mbar. The $WSe_2$/EG sample was then imaged using an Omicron variable temperature scanning tunneling microscope (STM) without any thermal treatment. The STM images were obtained at room temperature and in the constant-current mode, with an etched tungsten tip. The same system is equipped with a monochromatic Al-Kα source (E =1486.7 eV) and an Omicron Argus detector operating with pass energy of 15 eV. The spot size used during the acquisition is equal to 0.5 mm. Core-level spectra taken with 15 sweeps are analyzed with the spectral analysis software analyzer. (See: http://rdataa.com/aanalyzer).

[a]Department of Materials Science and Engineering and Center for 2-Dimensional and Layered Materials, The Pennsylvania State University, University Park, Pennsylvania, 16802, United States; [b]Department of Electrical Engineering, The Pennsylvania State University, University Park, Pennsylvania, 16802, United States; [c]Department of Materials Science and Engineering, The University of Texas at Dallas, Richardson, Texas 75080 United States; [d]Institute of Atomic and Molecular Sciences, Academia Sinica, Taipei 10617, Taiwan; [e]Physical Science and Engineering Division, King Abdullah University of Science and Technology, Thuwal, 23955-6900, Saudi Arabia. *email: jrobinson@psu.edu

**Heterostructure Synthesis on Epitaxial Graphene Substrates.**

Molybdenum disulfide synthesis on WSe$_2$/EG (Fig. S1a), often results in multilayer growth along edges and defects in WSe$_2$ due to higher reactivity at these sites. Cross-sectional HRTEM images of the MoS$_2$/WSe$_2$ heterostructure (Fig. S1b) scanned from center to edge of the heterostructures verifies the increase in top layer thickness. The WSe$_2$/MoSe$_2$/EG heterostructures cannot be identified in AFM morphology without the assistance of CAFM (Figs. S1c,d). Bilayer WSe$_2$ can also form on MoSe$_2$/EG. The positions of stacking trilayer are visualized in CAFM under $V_{bias}$ due to different electrical properties between WSe$_2$/EG and WSe$_2$/MoSe$_2$/EG (Fig. S1d). Electron energy loss spectra (EELS) and energy dispersive x-ray spectroscopy (EDS) data (Figs. S2a,b) verifies the heterostructure does not alloy, but are instead unique layers with pristine interfaces. In the case of MoS$_2$-WSe$_2$-EG, a multilayer region of MoS$_2$-WSe$_2$ was focused to ensure pristine layer formation beyond the monolayer structure, however, all electrical characterization presented later is on monolayer heterostructures. The WSe$_2$-MoSe$_2$-EG ordering is confirmed by comparing the intensity with that of bilayer-WSe$_2$-EG due to the similar atomic number between W and Mo atom (Fig. S2c).

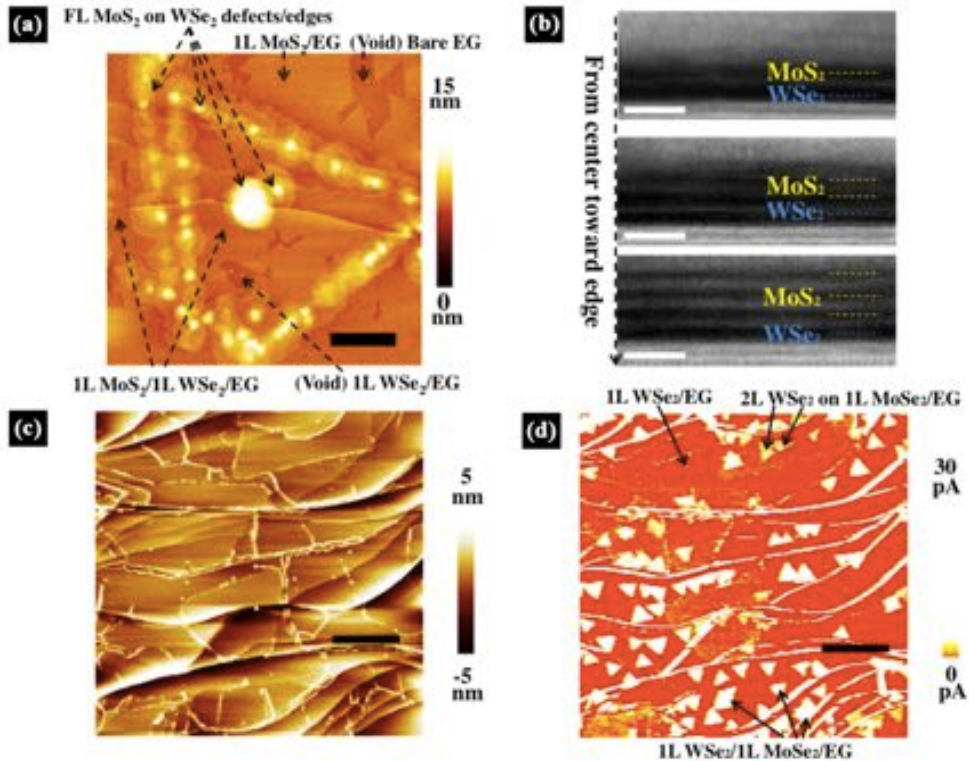

**Figure S1: MoS$_2$ crystals on WSe$_2$-EG and EG and WSe$_2$ crystals on MoSe$_2$-EG and EG:** (a) The MoS$_2$ crystals cover both of EG and WSe$_2$/EG after the CVD growth (Scar bar: 400 nm) (b) shows TEM profiles across the MoS$_2$-WSe$_2$-EG: Most of area in WSe$_2$ has MoS$_2$ in 1L thickness while the area surrounding the edge has multilayered MoS$_2$ potnetially due to higher reactivity around edges and defects, as shown in (a). (Scale bar in all the TEM image: 2 nm). (c,d) WSe$_2$ can keep growing on 1L WSe$_2$-1L MoSe$_2$-EG and evatually formed bilayer and above WSe$_2$ on MoSe$_2$-EG (c), which reduces the amplitude of tunnel current on the trilayered junciton. (Scar bar in (c,d): 1μm; The $V_{bias}$ in (d) is + 0.8 V)

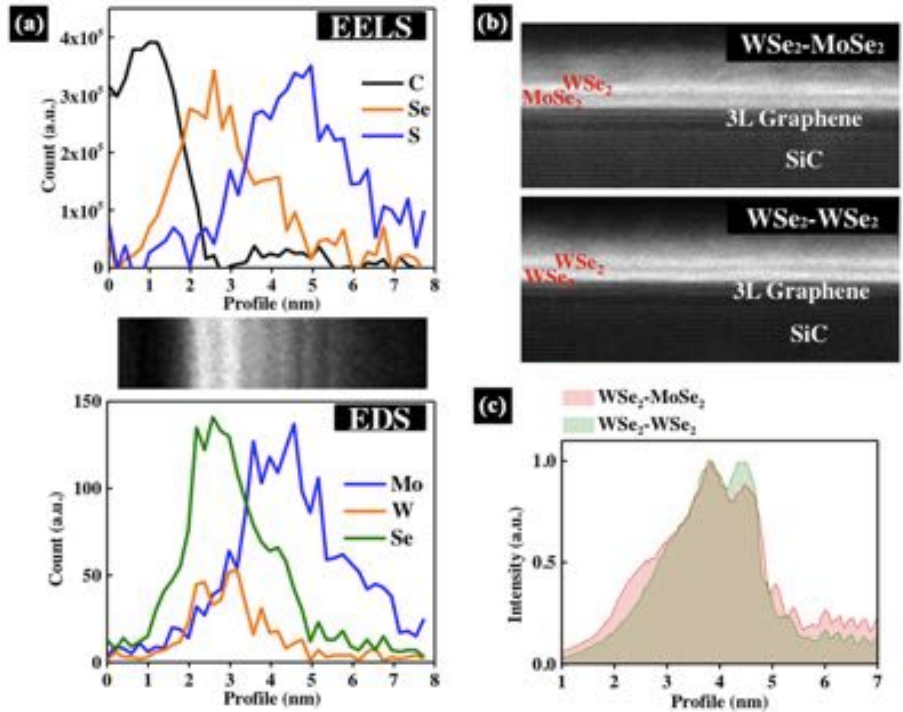

**Figure S2.** (a) The EELS and EDS cross-profiles in HAADF HRTEM of double junctions elemental distributions of $MoS_2$, $WSe_2$, and graphene; and indicates no alloys in each TMD layer. (b,c) The $WSe_2$-$MoSe_2$ area has significant contrast lost comparing to $WSe_2$-$WSe_2$ area (top and bottom HAADF), due to $MoSe_2$ layer.

**Raman, PL, and XPS of the Heterostructures:** Raman spectroscopy (Figs. S3a,b) confirms presence of significant fractions of monolayer $WSe_2$ ($E_{2g}$/$A_{1g}$ at 250 cm$^{-1}$ and 2LA at 263 cm$^{-1}$)[1] and $MoS_2$ ($E_{2g}$ at 383 cm$^{-1}$ and $A_{1g}$ at 404 cm$^{-1}$),[2] as well as monolayer $MoSe_2$ ($A_{1g}$ at 240 cm$^{-1}$ and $E^1_{2g}$ at 284 cm$^{-1}$) and indicates the absence of alloying. The spectroscopic signatures due to the interlayer coupling of $MoS_2$/$WSe_2$ are located at 285 cm$^{-1}$ ($E"$ for $MoS_2$)[3] and 309 cm$^{-1}$ (out-of-plane mode $A^2_{1g}$ for $WSe_2$)[3] (marked with asterisks in Fig. S3a). The normalized PL (Figs. S3c,d) under two different excited wavelengths also provide evidence that no alloying has occurred.

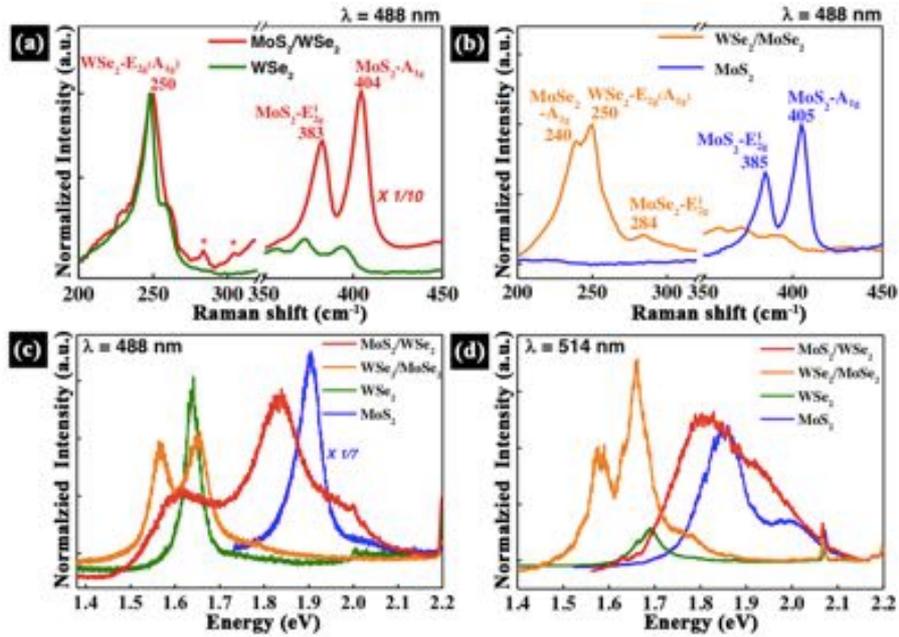

**Figure S3: The Raman PL spectra of heterostructures** (a) The spectrums clearly displays distinct features from $MoS_2/WSe_2$, (b) and $WSe_2/MoSe_2$ and indicate no alloy-like features. The asterisks indicate the signatures of their strong couplings and can also be found in mechanically stacked $MoS_2/WSe_2$.[3] (c) and (d) show the PL of heterostructures with intensity normalized to the Raman feature of SiC substrate with 488 and 514 nm excited laser, respectively. The peaks features in terms of signal intensity and peak shapes have been significant modulated versus the peaks of $MoS_2/EG$ and $WSe_2/EG$.

Prior to the TMD growth, XPS analysis was performed on epitaxial graphene (EG) synthetized on 6H-SiC(0001).[2] The C $1s$ core binding energy level was determined to be 284.1 eV and identical to the binding energy measured on a reference sample of HOPG. The core level spectra obtained after the growth of $MoS_2$ or $WS_2$ on EG are presented in Fig. S3 and compared in Table S1. The C $1s$ shifts by 0.3 eV to a higher binding energy after the formation of the first interface ($MoS_2/EG$ or $WSe_2/EG$) indicating p-doped graphene. The C $1s$ position remains the same after formation of the second interface in the trilayer heterostructure ($MoS_2/WSe_2/EG$ or $WSe_2/MoSe_2/EG$).

For comparison, a n-type $MoS_2$ bulk crystal indicates that the core level of Mo $3d_{5/2}$ (S $2p_{3/2}$) is located at 229.9 eV (162.7 eV),[4] The W $4f_{7/2}$ (Se $3d_{5/2}$) core level measured on a p-type $WSe_2$ crystal is located at 32.4 eV (54.9 eV). This suggests that the $MoS_2$ film exhibits a n-type conductivity and the $WSe_2$ layer shows a p-type behavior. We explain the difference between the n-$WSe_2$ (STS) vs. p-$WSe_2$ (XPS) by the difference in the local measurement for STS ($\leq 1nm^2$) vs. the surface sampling from the 0.5 mm diameter spot size used during the XPS acquisition. The photoemission also indicates the absence of any interaction between the two transition metal dichalcogenides or carbide formation with the underlying substrate. Mo-O bond formation is the only oxide detected at 236.7 eV (Mo $3d_{3/2}$). W-O bond formation, if present, is below the limit of detection as the W $4f$ oxidized state overlaps with the broad W $5p$ peak located at 37.6 eV. The formation of $WO_3$ oxide is frequent occurrence during the growth, with a peak position of W $4f_{5/2}$ at 37.5 eV.[5] Moreover, the XPS analysis shows the absence of any detectable Se-O and S-O oxides. Even the Mo-O bond is below the limit of detection after a thermal treatment in UHV at 250 °C. Table S1 provides the core level energies of $WSe_2$ and $MoS_2$ interfacing with graphene, which nearly the same position as the bulk samples p-$WSe_2$ and n-$MoS_2$. Interestingly, the doping level for $WSe_2$ in combination with graphene and $MoS_2$, is different in comparison to only $WSe_2$/graphene or $WSe_2$/$MoSe_2$. Also, $MoS_2$ is more n-type in a $MoS_2$/graphene than $MoS_2$/$WSe_2$.

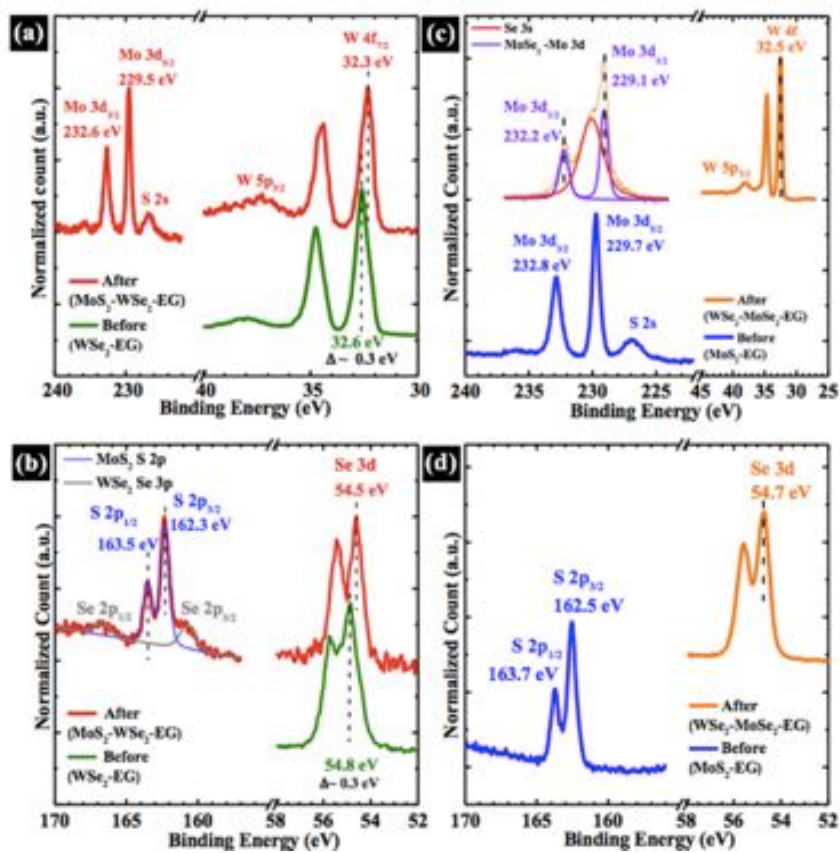

**Figure S4: XPS core shell analysis of (a,b) $MoS_2$/$WSe_2$/EG and (c,d) $WSe_2$/$MoSe_2$/EG heterostructures.** The photoemission indicates the absence of any interaction between the two TMDs or detectable TM-carbide formation with the underlying graphene/SiC. All oxides (W-O, Mo-O, Se-O, and S-O) are below the detection limit. In (c) the S $2s$ intensity associated with $MoS_2$/EG is also below the limit of detection after the $WSe_2$ growth, corresponding to the complete selenization of the $MoS_2$. Comparison of W 4f and Se 3d core shell doublet from $WSe_2$/EG and $MoS_2$/$WSe_2$/EG (a,b) showing an energy shift of ~0.3 eV, indicating $MoSe_2$ withdraws the negative charge to the bottom $WSe_2$.[6,7]

**Table S1. XPS core level position measured for different interfaces**

* The C 1s peak was convoluted to 3 components (carbide + graphene + buffer layer), the table shows only the C 1s peak from graphene.

|  | C $1s^{*}$ | W $4f_{7/2}$ | Se $3d_{5/2}$ | Mo $3d_{5/2}$ | S $2p_{3/2}$ |
|---|---|---|---|---|---|
| EG | 284.1 | - | - | - | - |
| WSe$_2$/EG | 284.4 | 32.6 | 54.8 | - | - |
| MoS$_2$/EG | 284.4 | - | - | 229.7 | 162.5 |
| MoS$_2$/WSe$_2$/EG | 284.3 | 32.3 | 54.5 | 229.5 | 162.3 |
| WSe$_2$/MoSe$_2$/EG | 284.4 | 32.5 | 54.7 | 229.1 | - |
| **Bulk samples** |  |  |  |  |  |
| n-MoS$_2$[4] | - | - | - | 229.9 | 162.7 |
| p-MoS$_2$[4] | - | - | - | 229.1 | 161.9 |
| p-WSe$_2$[8] | - | 32.7 | 54.9 | - | - |

**Heterostructure Synthesis on Traditional Substrates.**

Following the same processes described in the main text, we attempted to grow a vertical WSe$_2$/MoS$_2$ hetero-junction on a sapphire substrate. The detailed growth processes are described in the Method section of the main text. Fig. S4 (a) shows an optical micrograph after the CVD process, where the pre-growth WSe$_2$ is marked with black dashed line. In some cases, we observed that MoS$_2$ grew from the edge rather than on top of the WSe$_2$, which is clearly shown in Fig. S5 (a). The AFM image of Fig. S5 (b) confirms that the MoS$_2$ grows from the edge. We also find that there might have been some structural damage on pre-growth WSe$_2$ as found in Fig. S5 (b). Figs. S5 (c) and (d) show the Raman and PL spectrum of the WSe$_2$ before (black) and after (blue) synthesis of MoS$_2$/WSe$_2$/sapphire compared to pure MoS$_2$ (red), where the locations are indicated in Fig. S5 (a). From the Raman spectra, we find that the WSe$_2$ is replaced by sulfur and molybdenum to form WS$_2$ and MoS$_2$. The PL of WSe$_2$ is dramatically reduced and a signature of MoS$_2$ is found in the inner flake regime, which further suggests that the WSe$_2$ has been damaged and/or replaced during CVD process. The replacement of WSe$_2$ by sulfur and molybdenum on sapphire substrates may begin at defect sites within the flake or the chosen conditions for synthesis TMDs on graphene and sapphire, which needs further investigation.

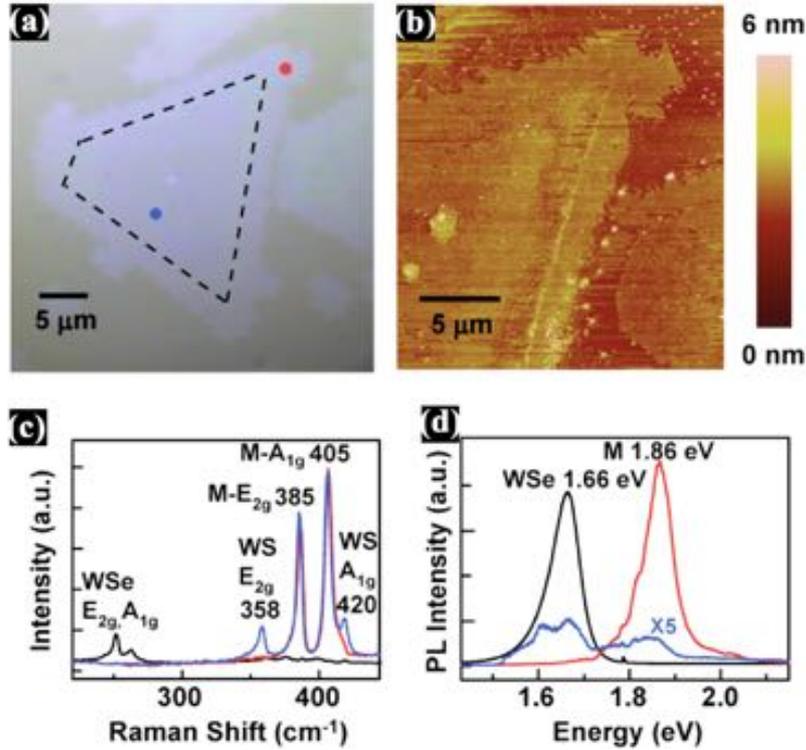

**Figure S5. The growth of vertical heterostructures on sapphire ends up with 2D alloys**. (a) shows the optical micrograph after CVD process, and the boundary of per-growth WSe$_2$ part is located with black dashed line. In some case, we observed that MoS$_2$ is growth from the edge instead on the top of the WSe$_2$, which is clearly shown in (a). The AFM image of (b) confirms that the MoS$_2$ grows from the edge. Some structural damages on per-growth WSe$_2$ are found in (b). (c,d) show the Raman and PL spectrum of the WSe$_2$ before (black) and after (blue) MoS$_2$ synthesis compared to bare MoS$_2$ (red), the examined positions are indicated in (a).

**Scanning Tunneling Spectroscopy: Error estimations for Band gap, VBM and CBM.**

In order to determine the electronic band gap value obtained for 1L WSe$_2$ and 1L MoS$_2$/1L WSe$_2$ the dI/dV curves in the STS were differentiated from an average of at least ten I-V curves acquired sequentially at a fixed position. The data were acquired close to the center of the heterostructure domains in order to avoid probing edge states along TMD edges which overlap with graphene.[9] The bandgap, the CBM and the VBM value and error given in the main text were estimated from several measurements. The CBM and VBM were acquired from the points where fitting lines intersect the both ends of bandgap region (where dI/dV curves present a parallel line). The bandgap is then determined by: [CBM$_{average}$ – VBM$_{average}$]. As an example, Fig. S6 shows the WSe$_2$ bandgap value extrapolated from 6 different dI/dV spectra. The mean bandgap is estimated at 1.83 eV with a standard deviation equal to 0.07, i.e. E$_g$(WSe$_2$) = 1.83±0.07 eV.

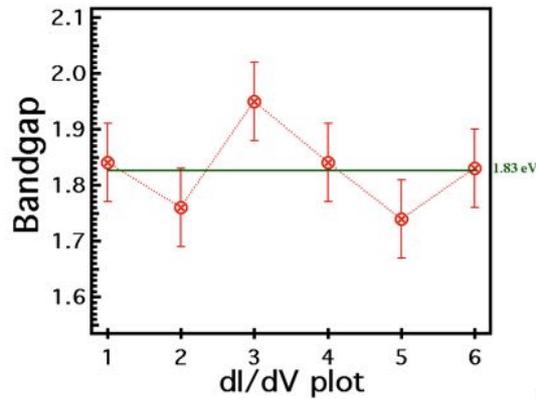

**Figure S6:** WSe$_2$ band gap values extrapolated from six dI/dV curves.

## Additional I-V plots from the CAFM measurement.

The measured spots were confirmed as the double junctions in the topography/CAFM images before the I-V curve measurement was accomplished. The AFM tip-loading force was increasing up to 5 nN, after which the curves are virtually identical due to the optimized contact area between tip and the samples as well as the removal of possible air/water layer in between.[10] Each plot in Fig. S7 is after the average of 3 repeated measurements on the same spot. The corresponding peak-to-valley current ratio (PVCR) and the voltage of the peak maximum are labeled. The variation in the peak position and width is likely related to defect formation within the layers and the interface of the layers.

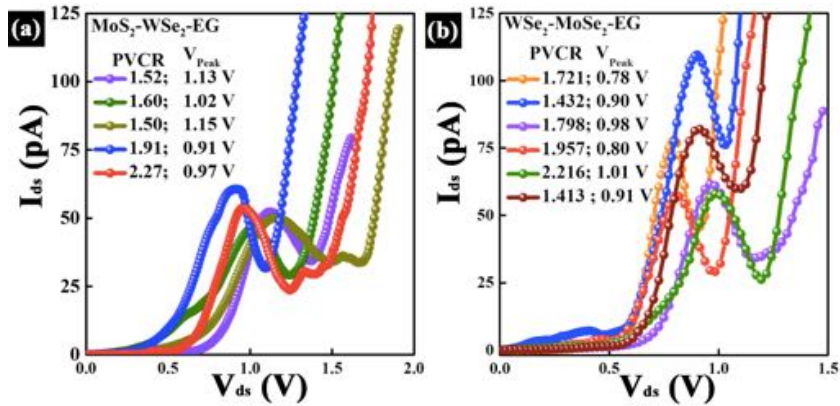

**Figure S7: Additional I-V plots.** The additional NDR curves from MoS$_2$-WSe$_2$-EG and WSe$_2$-MoSe$_2$-EG heterostructures are plotted in (a) and (b), respectively.

## Negative Differential Resistance in 2D heterostructures.

Resonant tunneling between two spatially separated quantum states can be used to realize negative differential conductance. Negative differential conductance holds the key for novel nano-electronic design options utilizing bistability and positive feedback. Novel memories, multi-valued logic and inductor-free compact oscillators and other electronic applications can benefit from a low power, low voltage negative differential conductance device. The resonant tunneling diode (RTD) has been a subject of intense study and design optimization, in Silicon Germanium and III-V heterostructure material systems for many years now. While theoretically capable of operating in extremely narrow voltage windows, the negative differential conductance of a resonant tunneling device, particularly at room temperature, is limited by scattering mechanisms, related to interfacial imperfections, which are unavoidable even when utilizing high vacuum advanced epitaxial growth technique. The interface related scattering reduces the sensitivity of resonant tunneling to an external bias, thereby increasing the voltage window over which negative differential conductance regime is observed.

van der Waals epitaxy of 2D materials can mitigate these issues and provide a materials platform for device engineers to obtain energetically sharp NDR features at room temperature leading to novel low power quantum tunneling devices. Absence of dangling bond in the interface and reduced interface roughness scattering in these two dimensional materials based devices makes it possible to obtain sharp NDR with very low full width half max voltage and relatively high PVCR in room temperature (Table S2 and Fig. S8).

Table S2: The table of comparison with other reported NDR

| System | $V_{peak}$ (V) | $J_{peak}$ ($\mu A/\mu m^2$) | $J_{valley}$ ($\mu A/\mu m^2$) | PVCR |
|---|---|---|---|---|
| $MoS_2/WSe_2$/EG (This work) | 0.8 to 1.1 | 0.15 | 0.075 | 1.5 to 2.3 (T=300K) |
| $WSe_2/MoSe_2$/EG (This work) | 0.7 to 1.1 | 0.12 | 0.060 | 1.4 to 2.2 (T=300K) |
| Gr/BN/Gr [11] | 0.8 | 0.22 | 0.13 | 1.7 (T=300K, no gating) |
| 3-layered $MoS_2$ [12] | 0.51 | 3 | 2.7 | 1.1 (T=60K, no gating) |
| Si/SiGe [13] | 0.22 | 3 | 0.83 | 3.6 (T=300K) |

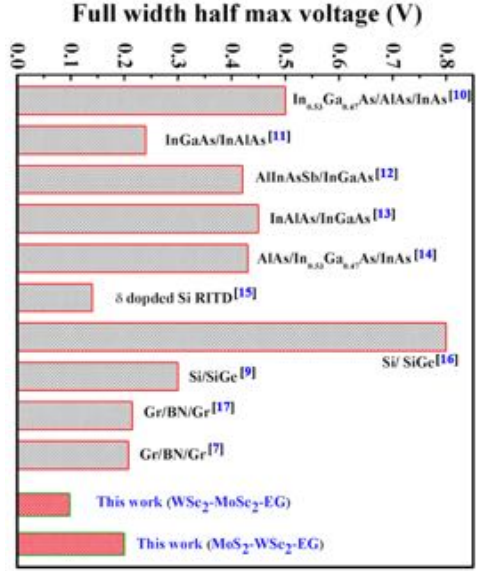

**Figure S8.** Comparison of Full Width Half Max voltage with other reported results in room temperature.[11,13–21]

## DFT+NEGF calculation.

To make a direct comparison between the measurements and the quantum transport in these vertical resonant tunneling devices, we carry out non-equilibrium quantum transport calculations by using density functional theory (DFT) coupled with the Non-Equilibrium Green's function (NEGF) formalism.[22]

**Electronic properties of individual layers:** We carry out the DFT simulations by using *QuantumWise simulator*, Atomistix Toolkit (ATK), (www.quantumwise.com).[23] The electronic properties of different crystals are calculated by using generalized gradient approximations (GGA). Within the DFT, the valence band wave functions of different atoms are treated in terms of a linear combination of atomic orbitals (LCAO) and the electronic properties of core electrons is described by norm-conserving Troullier-Martins pseudopotentials. In the LCAO pseudopotential calculations we consider the Perdew-Burke-Ernzerhof (PBE) approximation for the exchange-correlation functional along with double $\zeta$-polarized (DZP) basis on the atoms. These electron wave functions are usually comparable to well converged plane wave basis sets. To incorporate the long range van der Waals correction in interlayer interaction within the GGA approximation, we have included Grimme's DFT-D2 functional with S6 = 0.75 under PBE functional. Moreover, to calculate the individual band structure properties we use a k-point sampling of 5x5x1 in the Brillouin zone. The tolerance parameter was set to a value of $10^{-5}$ with maximum steps of 200, and a Pulay mixer algorithm was used as the iteration control parameter with mesh cut-off energy of 150 Ry on a real space grid of charge density and potentials. The lattice parameters of individual crystals are listed in Table S-II and the monolayer bandstructures are shown in Fig. S9.

**Table S3: Lattice constant of monolayer TMD used in DFT-GGA calculations**

|   | $MoS_2$ | $WSe_2$ | $MoSe_2$ | EG |
|---|---|---|---|---|
| *a* | 3.17 Å | 3.288 Å | 3.29 Å | 2.461 Å |

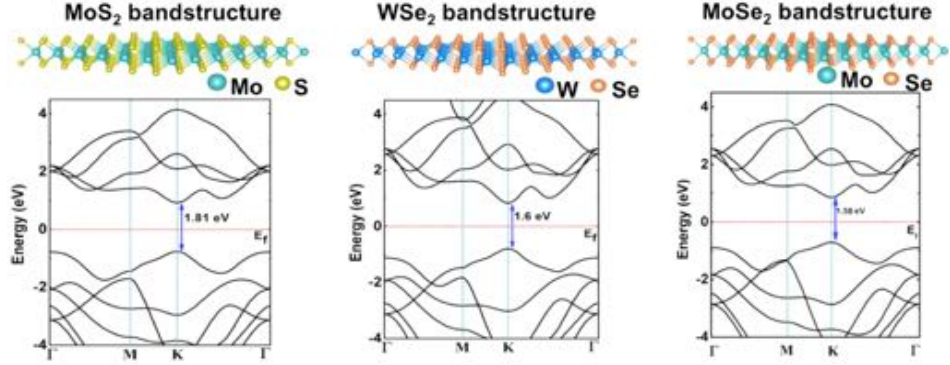

**Figure S9:** Electronic band structure of individual band structure of different monolayer TMD and their corresponding band gap by DFT-GGA calculations. The band gap gives very good agreement with the existing experimental results.[24]

**van der Waals Heterostructure device simulation setup:** The device configuration of our simulation is shown in Fig. 3b in the article. To construct the device configuration, we first utilize a supercell of $MoS_2$ ($MoSe_2$) /$WSe_2$ /4-layer EG that is periodic in the $x$ and $y$ directions. To construct the $MoS_2$ /$WSe_2$ /EG (or $WSe_2$ /$MoSe_2$ /EG) interface (here we consider $0^o$ rotation between the TMDs to minimize the number of atoms to reduce the computational burden), we merge 3x3 supercell of monolayer TMDs with 4x4 graphene to reduce the lattice mismatch between the di-chalcogenides and graphene hexagonal unit cell. The spin-orbit coupling is not considered here to reduce the computational complexity in our device simulations. We optimize the structure by using Quasi-Newton method until all the forces acting on atoms become smaller than 0.02 eV/Å. The structure optimization provides a van der Waals distance of 3.53 Å between $WSe_2$ and graphene and 3.41Å between $MoS_2$ ($MoSe_2$) and $WSe_2$. We place this optimized supercell on Pt (111) surface (applying the strain on Pt only) and perform the structure optimization again by keeping the lattice parameters fixed, so that the effect of extra strain due to Pt is restricted to the Pt(111) surface. Thus, the active region of the device as shown in Fig. S10 used for quantum transport simulation by the NEGF-DFT code consists of nine Pt(111) layers on the left, monolayer of $MoS_2$, monolayer of $WSe_2$ and 4 layers of EG on the right. This active region is then attached to two semi-infinite ideal electrodes (left and right) where left is the Pt(111) electrode and right is the Graphene electrode. We want to mention that, here we consider 4 layers of EG instead of 3 as observed in the experimental setup. For the self-consistent NEGF simulations, the Brillouin zone of the superlattice was sampled by $5 \times 5 \times 101$ k-point grid along with a mesh cutoff energy of 150 Ry and with the same parameters as used before to calculate the electronic bandstructure. These numerical parameters are quite sufficient to attain a total energy convergence of 0.01 meV/unit cell within the self-consistent loop of the simulation of the device.

Under steady state transport, the NEGF formalism works with two principal quantities, the retarded Green's function $G^R(E)$, and the lesser Green's function $G^<(E)$. These define the accessible quantum states of the electrons and their occupancy into those states, respectively. In this formalism, the key quantity is to compute the Hamiltonian ($H$) and the overlap matrix ($S$) from the self-consistent DFT loop of the density matrix $\rho = (1/2\pi i)\int G^<(E)dE$.

Under local orbital basis set $\{\psi_i\}$, the Hamiltonian is defined as $H_{ij} = \langle \psi_i | \hat{H}_{KS} | \psi_j \rangle$ where $\hat{H}_{KS}$ is the Kohn-Sham Hamiltonian as obtained from the self-consistent DFT, whereas the overlap matrix is defined as $S_{ij} = \langle \psi_i | \psi_j \rangle$. From the knowledge of this $H$ and $S$ matrices along with the energy Eigen value matrix $E$ of the system one can construct the retarded Green's function, which is the only requisite to for the calculation of coherent quantum transport. Thus, the retarded Green's function can be defined as

$$G^R_{k_P} = \left[ ES - H - \Sigma_{L,k_P} - \Sigma_{R,k_P} \right]^{-1} \tag{S1}$$

where $\Sigma_{L,k_P}$ and $\Sigma_{R,k_P}$ are the self-energy matrices of the left and right electrode contacts, respectively. Using this, the transmission function can be easily given by

$$T(E,V,k_P) = Trace[\Gamma_{R,k_P}(E) G^R_{k_P}(E) \Gamma_{L,k_P}(E) (G^R_{k_P}(E))^\dagger] \tag{S2}$$

in which $\Gamma_{L,k_P}$ and $\Gamma_{R,k_P}$ are the level broadening matrices $\left[ \Gamma_{L/R,k_P}(E) = (i/2)\{\Sigma_{L/R,k_P}(E) - \Sigma^\dagger_{L/R,k_P}(E)\} \right]$ associated with the left and right electrodes, respectively, which are the anti-Hermitian components of the self-energies of the semi-infinite ideal electrodes. It can be then easily computed the current from left to right electrode as

$$I(V_{ds}) = \frac{2q}{h} \int_{BZ} dk_{||} \int dE\ T(E, k_{||}, V_{ds}) \left[ f\left(\frac{E-E_{f_1}}{k_B T}\right) - f\left(\frac{E-E_{f_2}}{k_B T}\right) \right] \tag{S3}$$

where we assume the electrodes are infinite reservoirs characterized by the Fermi function $f(E - \mu_{L/R})$ and the difference between the chemical potentials $E_{f1} - E_{f2} = qV_{ds}$ defines the applied bias for the self-consistent non-equilibrium transport. Here we want to note that, in our device simulations we use the applied bias at the left electrode by keeping the right electrode bias at zero. Finally the total current is expressed as the sum of both of the spin up and spins down components of the current. Further, to calculate the local density of states (LDOS) we use the formulation,

$$D(E,r) = \sum_{i,j} \rho_{i,j}(E) \psi_i(r) \psi_j(r) \tag{S4}$$

where $\rho(E)$ is the spectral density matrix.

To provide insight on the observed resonant tunneling, we calculate the transmission Eigen states in the three resonance peaks (P1, P2 and P3 in Fig. 3d). Inspection of the localized molecular orbitals of the transmission Eigen states in Fig. S10 reveals that all three resonance peaks in Fig.3d originate from a combination of the *Pt* (*s*-orbital), WSe$_2$ (*p*-orbital of Se, and $p_z$, $d_{z^2}$-orbital of W) and graphene layers (*p*-orbital). Interestingly, in the case of MoS$_2$-WSe$_2$-EG, the MoS$_2$ and the top graphene layer (Gr1 in Fig. S10) do not contribute to the strong transmission peaks.

Zhu et al.,[25] recently showed that the in-plane $d_{x2-y2} + d_{xy}$ orbitals of the transition metal are mainly responsible for the spin splitting and in the out-of-plane orientation the chalcogen $p_z$ and transition metal $d_{z2}$ states play no role for spin splitting arising from spin orbit coupling. Therefore, for the out-of plane vertical tunnel transport in the heterostructure as discussed here, we can argue that the spin-orbit coupling will have negligible effect on transport mechanism. To understand whether this orbital contribution of $WSe_2$ comes from its electrons or holes, we further study the position and energy dependent local density of states (LDOS), (N (z, E), eq. S4) of this nanoscale device at the peak voltage of +1.0 V (Fig. S11a) and to correlate the energy positions of the resonance peaks (peaks P1, P2, P3 in Fig. 3d) that occur within this Fermi window of $E_{f1} - E_{f2}$ = 1.0 V.

Now, with the application of positive bias, the Fermi level in the *Pt* electrode is lowered with respect to the Fermi level of the graphene electrode as shown in Fig. 3b. If we see the energy locations of the corresponding transmission peaks in Fig. 3d and map with the energy positions in the LDOS in Fig. S11a we find that the hole energy states in the $WSe_2$ layer contribute to the transmission. Therefore, both the LDOS and transmission Eigen state indicates a strong coupling between the available states from the *Pt* electrons (or holes) with the $WSe_2$ holes and the graphene holes that provides resonant conduction states enclosed by the chemical potential of the *Pt* and the graphene electrodes. So, we conclude that the orbital contribution in transmission Eigen states of the resonant peaks presented in Fig. S10 are due to the confinement of holes in $WSe_2$ by the $MoS_2$ arising from valence band offset for the applied positive bias condition. The LDOS plot further points out that that interatomic electronic interaction between $MoS_2$ and $WSe_2$ makes them n- and p-type, respectively; which is in good agreement with our experiment results and other theoretical works.[26,27] Conversely, with the application of negative bias, the Fermi level in the *Pt* electrode will elevate with respect to the Fermi level of the graphene electrode. Therefore, in the negative bias regime, electronic states contribute in this tunneling process instead of hole states. Now, in this staggered gap (type II)[6,28] heterostructure, the conduction band of n-type $MoS_2$ remains lower than the conduction band of p-type $WSe_2$ and forms a conduction band offset in this device. Therefore, the confined electrons in $MoS_2$ due to this conduction band offset offers the necessary bound states of the resonant tunneling in the negative bias regime.

Further, Fig. S11b is the k-resolved transmission spectra within the hexagonal Brillouin zone of $MoS_2$-$WSe_2$-EG, which demonstrates the presence of "sharp" spikes at specific points in the Brillouin zone. This phenomenon corresponds to resonant tunneling due to the conservation of the momentum during this resonant tunneling process. In contrast, conventional tunneling would result in a diffuse transmission spectra over the entire Brillouin zone. A similar observation in terms of Eigen states, LDOS, and k-resolved transmission spectra is also realized in $WSe_2$-$MoSe_2$-EG heterostructure at its peak current bias point (Fig. S12 and its corresponding transmission peak labeled *P*, Fig. 3d).[29]

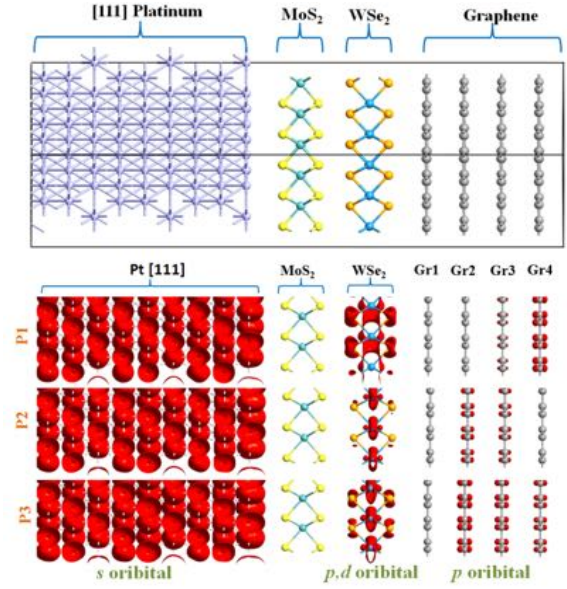

**Figure S10:** Transmission Eigen states contribute to the transmission in the three peaks (peak P1, P2 and P3 of Fig. 3d) of the transmission at an applied bias $V_d = +1.0V$ in the $MoS_2$-$WSe_2$-EG heterostructure.

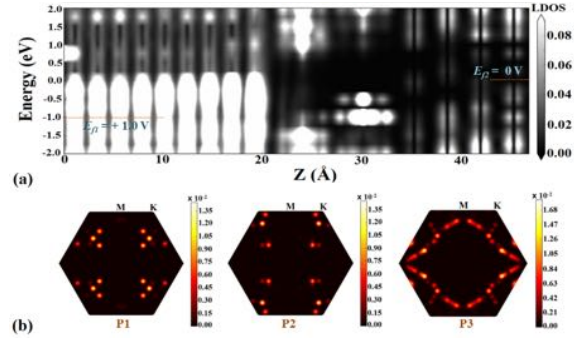

**Figure S11:** (a) local density of states (LDOS) at $V_d = +1.0$ V and (b) corresponding *k*-resolved contour plot of the transmission peaks (peak P1, P2 and P3 of Fig 3d) within the hexagonal Brillouin zone provide insight into the physical origin of the resonant tunneling.

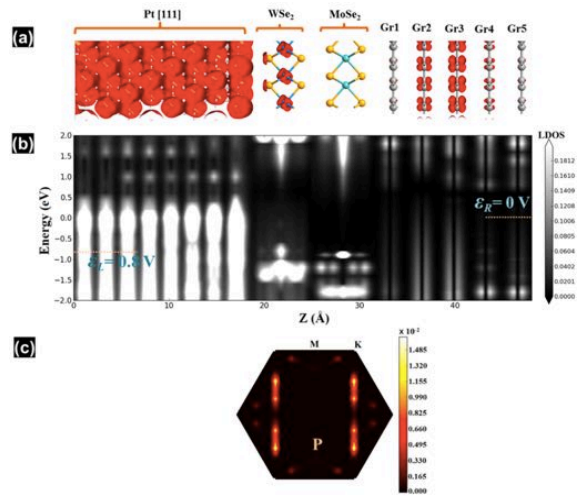

**Figure S12:** (a) Transmission Eigen states that contribute to the peak of the transmission at an applied bias V = 0.8 V in the WSe$_2$/MoSe$_2$/EG heterostructure and (b) local density of states (LDOS) at V = 0.8 V. (c) corresponding *k*-resolved contour plot of the transmission peak (peak P in Fig. 3d) within the hexagonal Brillouin zone provide insight into the physical origin of the resonant tunneling in WSe$_2$/MoSe$_2$/EG.

**Reference:**

1. Huang, J.-K. *et al.* Large-area synthesis of highly crystalline WSe$_2$ monolayers and device applications. *ACS Nano* **8,** 923–930 (2014).

2. Lin, Y.-C. *et al.* Direct Synthesis of Van der Waals Solids. *ACS Nano* **8,** 3715–3723 (2014).

3. Chiu, M.-H. *et al.* Spectroscopic signatures for interlayer coupling in MoS$_2$-WSe$_2$ van der Waals stacking. *ACS Nano* **8,** 9649–9656 (2014).

4. McDonnell, S., Addou, R., Buie, C., Wallace, R. M. & Hinkle, C. L. Defect-dominated doping and contact resistance in MoS$_2$. *ACS Nano* **8,** 2880–2888 (2014).

5. Wang, H. *et al.* MoSe$_2$ and WSe$_2$ nanofilms with vertically aligned molecular layers on curved and rough surfaces. *Nano Lett.* **13,** 3426–3433 (2013).

6. Fang, H. *et al.* Strong interlayer coupling in van der Waals heterostructures built from single-layer chalcogenides. *Proc. Natl. Acad. Sci. U. S. A.* **111,** 6198–6202 (2014).

7. Coy Diaz, H., Addou, R. & Batzill, M. Interface properties of CVD grown graphene transferred onto MoS2(0001). *Nanoscale* **6,** 1071–1078 (2014).

8. McDonnell, S. *et al.* Hole Contacts on Transition Metal Dichalcogenides: Interface Chemistry and Band Alignments. *ACS Nano* **8,** 6265–6272 (2014).

9. Zhang, C., Johnson, A., Hsu, C.-L., Li, L.-J. & Shih, C.-K. Direct imaging of band profile in single layer MoS$_2$ on graphite: quasiparticle energy gap, metallic edge states, and edge band bending. *Nano Lett.* **14,** 2443–2447 (2014).

10. Lee, G. H. *et al.* Electron tunneling through atomically flat and ultrathin hexagonal boron nitride. *Appl. Phys. Lett.* **99,** (2011).

11. Mishchenko, A. *et al.* Twist-controlled resonant tunnelling in graphene/boron nitride/graphene heterostructures. *Nat. Nanotechnol.* **9**, 808–813 (2014).

12. Nguyen, L.-N. *et al.* Resonant tunneling through discrete quantum states in stacked atomic-layered $MoS_2$. *Nano Lett.* **14,** 2381–2386 (2014).

13. Jin, N. *et al.* Diffusion Barrier Cladding in Si / SiGe Resonant Interband Tunneling Diodes and Their Patterned Growth on PMOS Source / Drain Regions. *IEEE Trans. Electron Devices* **50,** 1876–1884 (2003).

14. Smet, J. H., Broekaert, T. P. E. & Fonstad, C. G. Peak-to-valley current ratios as high as 50:1 at room temperature in pseudomorphic In0.53Ga0.47As/AlAs/InAs resonant tunneling diodes. *J. Appl. Phys.* **71,** 2475 (1992).

15. Day, D. J., Yang, R. Q., Lu, J. & Xu, J. M. Experimental demonstration of resonant interband tunnel diode with room temperature peak-to-valley current ratio over 100. *J. Appl. Phys.* **73,** 1542–1544 (1993).

16. Su, Y.-K. *et al.* Well width dependence for novel AlInAsSb/InGaAs double-barrier resonant tunneling diode. *Solid. State. Electron.* **46,** 1109–1111 (2002).

17. Tsai, H. H., Su, Y. K., Lin, H. H., Wang, R. L. & Lee, T. L. P-N Double Quantum Well Resonant Interband Tunneling Diode with Peak-to-Valley Current Ratio of 144 at Room Temperature. *IEEE Electron Device Lett.* **15,** 357–359 (1994).

18. Evers, N. *et al.* Thin Film Pseudomorphic AlAdIn0.53Ga0.47As/InAs Resonant Tunneling Diodes Integrated onto Si Substrates. *IEEE Electron Device Lett.* **17,** 443–445 (1996).

19. Rommel, S. L. *et al.* Epitaxially grown Si resonant interband tunnel diodes exhibiting high current densities. *IEEE Electron Device Lett.* **20,** 329–331 (1999).

20. See, P. *et al.* High performance Si/Si1-x/Gex resonant tunneling diodes. *IEEE Electron Device Lett.* **22,** 182–184 (2001).

21. Britnell, L. *et al.* Resonant tunnelling and negative differential conductance in graphene transistors. *Nat. Commun.* **4,** 1794 (2013).

22. Datta, S. *Quantum Transport-Atom to transistor*. (Cambridge University Press, 2005).

23. QuantumWise simulator, Atomistix ToolKit (ATK). at <www.quantumwise.com>

24. Ghosh, R. K. & Mahapatra, S. Monolayer Transition Metal Dichalcogenide Channel-Based Tunnel Transistor. *IEEE J. Electron Devices Soc.* **1,** 175–180 (2013).

25. Zhu, Z. Y., Cheng, Y. C. & Schwingenschl, U. Giant spin-orbit-induced spin splitting in two-dimensional transition-metal dichalcogenide semiconductors. *Phys. Rev. B* **84,** 153402 (2011).

26. Terrones, H., López-Urías, F. & Terrones, M. Novel hetero-layered materials with tunable direct band gaps by sandwiching different metal disulfides and diselenides. *Sci. Rep.* **3,** 1549 (2013).

27. Gong, C. *et al.* Band alignment of two-dimensional transition metal dichalcogenides: Application in tunnel field effect transistors. *Appl. Phys. Lett.* **103,** 053513 (2013).

28. Lee, C.-H. *et al.* Atomically thin p–n junctions with van der Waals heterointerfaces. *Nat. Nanotechnol.* **9,** 676–681 (2014).

29. MacLaren, J. M., Zhang, X.-G. & Butler, W. H. Validity of the Julliere model of spin-dependent tunneling. *Phys. Rev. B* **56,** 11827 (1997).